\documentclass[aps,superscriptaddress,pra,twocolumn,,showpacs]{revtex4-1}
\usepackage{amsfonts, amsmath, amssymb, bm, bbm, graphicx, epsfig, epstopdf}
\usepackage{braket}
\usepackage{dcolumn}
\usepackage{textcomp}
\usepackage{hyperref}
\usepackage{color}
\definecolor{forestgreen}{RGB}{34,139,34}

\usepackage[normalem]{ulem} 
\usepackage{color} 

\begin{document}
	\title{Protocols for long-distance quantum communication with single $^{167}$Er ions}
	
\author{F. Kimiaee Asadi}
\affiliation{Institute for Quantum Science and Technology, and Department of Physics \& Astronomy, University of Calgary, 2500 University Drive NW, Calgary, Alberta T2N 1N4, Canada}
\author{S. C. Wein}
\affiliation{Institute for Quantum Science and Technology, and Department of Physics \& Astronomy, University of Calgary, 2500 University Drive NW, Calgary, Alberta T2N 1N4, Canada}
\author{C. Simon}
\affiliation{Institute for Quantum Science and Technology, and Department of Physics \& Astronomy, University of Calgary, 2500 University Drive NW, Calgary, Alberta T2N 1N4, Canada}
\begin{abstract}
    We design a quantum repeater architecture using individual $^{167}$Er ions doped into Y$_2$SiO$_5$ crystal. This ion is a promising candidate for a repeater protocol because of its long hyperfine coherence time in addition to its ability to emit photons within the telecommunication wavelength range. To distribute entanglement over a long distance, we propose two different swapping gates between nearby ions using the exchange of virtual cavity photons and the electric dipole-dipole interaction. We analyze their expected performance, and discuss their strengths and weaknesses. Then, we show that a post-selection approach can be implemented to improve the gate fidelity of the virtual photon exchange scheme by monitoring cavity emission. Finally, we use our results for the swapping gates to estimate the end-to-end fidelity and distribution rate for the protocol.
\end{abstract}

	\maketitle
\section{Introduction}\label{ssec:intro}
Future quantum networks will require the ability of long-distance communication \cite{kimble2008quantum, christoph, wehner}. Although we have an existing global fiber optics network for classical communication, the bottleneck for long-distance quantum communication is the unavoidable transmission loss through fibers. Classical communication overcomes this problem by amplifying signals, however, due to the no cloning theorem the use of amplifiers is prohibited in quantum communication. Therefore, to circumvent this exponential decay of transmitted photons, the use of a quantum repeater has been suggested \cite{briegel, DLCZ}. Quantum repeaters aim to establish entanglement between two distant locations. Most of the quantum repeater protocols that have been proposed so far focus on atomic ensemble-based quantum memories and linear optics for
entanglement generation and distribution \cite{sangouard}.
This is an attractive route as it requires only a few relatively simple components. However, when using linear optics, the success probability for entanglement swapping cannot exceed $1/2$, resulting in relatively low entanglement distribution rates. Using single-emitter-based quantum repeater protocols, on the other hand, one can perform entanglement swapping with a higher success probability\cite{childress, sangouard2009, Er-Eu, santra, rozpkedek, kumar}.

Several works have demonstrated the ability to individually address single rare-earth (RE) ions \cite{utikal, yin2013, kolesov, zhong2018, dibos}.
RE ions in general have a smaller sensitivity to lattice phonons and experience little spectral diffusion \cite{spectral-diffusion} compared to quantum dots and NV centers in diamond.
In addition, most other quantum systems, require the use of microwave (MW) to optical transducers (e.g., superconducting qubits \cite{kumar}) or the frequency downconversion to telecommunication wavelength (e.g., defects  in  diamond and  quantum  dots \cite{NV, quantumdot}) to match the low-loss wavelength range of fibers. 
However, the erbium (Er) RE ion has a unique feature, which is its ability to emit photons in the conventional telecommunication wavelength window.  Moreover, significant enhancements of RE ion emission rates, including Er, have been demonstrated \cite{dibos, casabone, raha, Thiel, zhong2015}. 

In $^{168}$Er with zero nuclear spin, the relevant coherence time is that of the electronic spin. Therefore, until recently, one challenge for using an $^{168}$Er ion as a quantum memory was its short spin coherence time. For a single $^{168}$Er ion doped in yttrium orthosilicate Y$_2$SiO$_5$ crystal ($^{168}$Er:YSO) in the presence of a strong magnetic field, a spin coherence lifetime of a few milliseconds is expected in low temperatures, which is not quite long enough for a repeater protocol. Therefore, in our previous work, we proposed a quantum repeater architecture combining an individual $^{168}$Er ion and europium ($^{151}$Eu) RE ion, which serve as a spin-photon interface and long-term memory, respectively \cite{Er-Eu}.
In this scheme to perform a swapping gate using the electric dipole-dipole interaction, Er-Eu ions should be close-lying. Hence, fabricating and identifying suitable Er-Eu ion pairs is a main challenge of this scheme.

Recently, a hyperfine coherence time of 1.3 s has been measured for an ensemble of $^{167}$Er:YSO using a strong external magnetic field 
\cite{sellars}. Instead of applying a large magnetic field, it is also possible to extend the coherence time using the 
zero first-order Zeeman (ZEFOZ) technique. For the $^{167}$Er ion, transitions with ZEFOZ shift exist with and without the external magnetic field \cite{zefoz-Er}.
The long hyperfine coherence time of $^{167}$Er suggests that it could serve as both the spin-photon interface emitting telecom photons and the long-lived quantum memory needed to implement a repeater protocol. These advantages, in addition to the narrow optical transitions, have made $^{167}$Er:YSO a very promising material platform for quantum communication.

In this paper, we propose and analyze a scheme to design quantum repeaters using single $^{167}$Er ions. We consider individual $^{167}$Er ions doped into a high quality factor YSO photonic crystal cavity. The presence of the cavity improves the intrinsic low radiative decay rate of the Er ion, increases the single-photon indistinguishability, and enhances the collection of photons into the desired transmission channel.
We first explain how to generate entanglement between remote $^{167}$Er ions over elementary links. 
Entanglement swapping between two ions within each cavity is then performed to extend the range of entanglement to successively longer distances.
Building on earlier work, we propose two different schemes to perform the entanglement swapping step of the repeater protocol deterministically. In the first scheme the controlled interaction between ions is achieved by the exchange of virtual cavity photons. In the second scheme the interaction is mediated by the electric dipole-dipole interaction between the ions. We also propose a method to improve the fidelity of the first scheme at the cost of some efficiency by monitoring cavity emission in order to post-select successful gates. We then determine the fidelity of each swapping gate scheme and finally estimate the end-to-end fidelity of the proposed single Er repeater protocol.

The paper is organized as follows: In Sec \ref{ssec:proposal}, we introduce our quantum repeater protocol. Sections \ref{ssec:fidelity} and \ref{ssec:rate} deal with the estimation of the fidelity and efficiency, and the entanglement generation rate of the repeater protocol, respectively.
The implementation of the protocol as well as the advantages and disadvantages of each of the entanglement swapping schemes are discussed in Sec.\ref{ssec:implementation}. 
We conclude with future directions in Sec. \ref{ssec:conclusion}

\section{Proposal}\label{ssec:proposal}

Each node consists of an optical cavity fabricated in the YSO host crystal that is doped with a pair of $^{167}$Er ions. 

\begin{figure}
	\includegraphics[width=8.5cm]{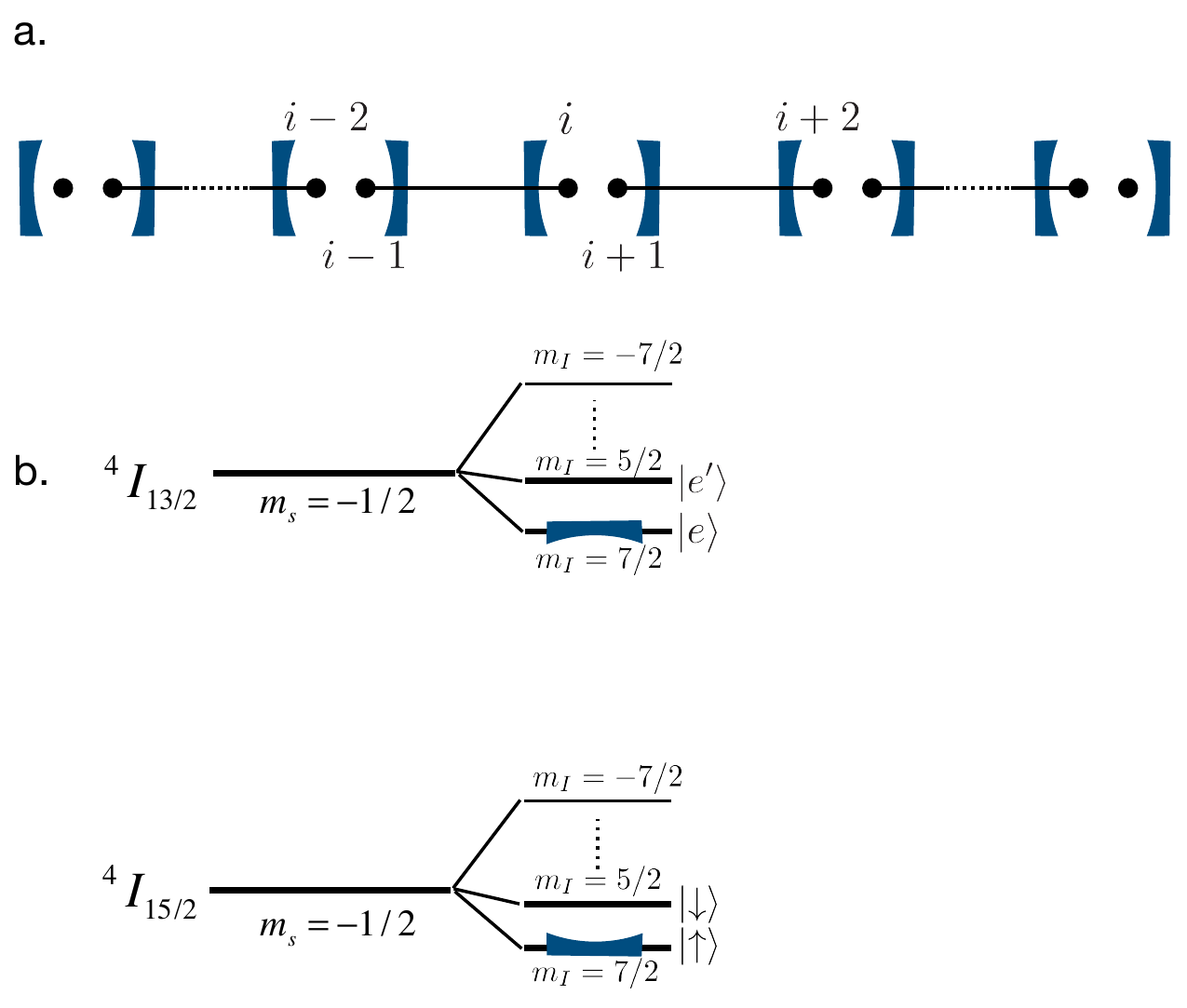}
	\caption{\textbf{a.} In each cavity there is a pair of $^{167}$Er ions (black circles). Black lines represent entanglement between degenerate ions over elementary links. \textbf{b.} Energy level structure of the ion. In each ion the $\ket{\uparrow}$--$\ket{e}$ transition is coupled to the cavity.}\label{fig:main}
\end{figure}

In the presence of a strong magnetic field along the $D_1$ axis and temperature of $1.6$ K or less, the ground state electron spin freezes at the lower level.
In our scheme, the $m_{I}=\frac{7}{2}$ and $m_{I}=\frac{5}{2}$ hyperfine states of the lowest spin state are used as qubit states $\ket{\uparrow}$ and $\ket{\downarrow}$, respectively, as shown in Fig\ref{fig:main}.b.
The oscillator strength for $\Delta m_I=-1 (+1)$ transitions relative to the $\Delta m_I=0$ is about 2.5\% (3.1\%) for transitions involving the $m_{I}=\frac{7}{2}$ hyperfine state \cite{sellars}. Therefore, with the use of a cavity, it is possible to utilize an L-type system, where the excited state has a high probability to decay to the initial ground state. This probability can be further increased by using a resonant cavity and therefore, we can ignore the other weak transitions.

 \subsection{Entanglement Generation} \label{ssec:entanglement}
 
To generate entanglement over an elementary link between Er ions in remote cavities, 
such as $Er_i$ and $Er_{i-1}$, we follow the same scheme as \cite{barrett, Hanson}. Initially, the $\ket{\uparrow}$--$\ket{e}$ transition of each ion is coupled to its respective cavity. First, both ions are optically pumped into the $m_{I}=\frac{7}{2}$ hyperfine ground state. 

Using optical Raman pulses, each ion is then prepared in the superposition of $\ket{\uparrow}$ and $\ket{\downarrow}$ states. Ions are then excited to the $\ket{e}$ state using a short laser pulse resonant with the $\ket{\uparrow}$--$\ket{e}$ transition. After sufficient time has passed to allow a possible photon to be emitted through the cavity mode, optical Raman pulses are applied to flip the qubit state. This is followed by another optical excitation to the $\ket{e}$ state to emit a possible photon. The second round of excitation is key to overcoming infidelity caused by photon loss in the fiber in the event that both ions emit a photon. The emitted photons are then collected and transmitted to a beam splitter located half-way in between the ions. The detection of two consecutive single photons will then leave remote ions in an entangled Bell state  \cite{conditional}
\begin{equation}
\label{eq:Bellstate}
 \ket{\psi^{\pm}}_{Er_i,Er_{i-1}}=\frac{1}{\sqrt{2}}(\ket{\uparrow\downarrow}\pm \ket{\uparrow \downarrow}).
 \end{equation}
Here the sign + (-) depends on whether the same (different) detectors detect photons.

\subsection{Entanglement swapping}
After generating entanglement over elementary links, entanglement is swapped between nearby ions within each cavity (e.g.,  Er$_i$ and Er$_{i+1}$ in Fig. \ref{fig:main}.a).  
This can be done by performing a CNOT gate between the ions and then measuring the control (target) ion in the X (Z) basis. Measurement in the Z basis is achieved by the optical excitation of ions from the ground state $\ket{\uparrow}$ to the excited state $\ket{e}$ while this transition is coupled to the cavity. To perform the spin readout in X basis, we need to coherently rotate the ion (to make $\ket{\downarrow}\rightarrow 1/\sqrt{2} (\ket{\downarrow}-\ket{\uparrow})$ and $\ket{\uparrow}\rightarrow 1/\sqrt{2} (\ket{\uparrow}+\ket{\downarrow})$) followed by a measurement in the Z basis. Depending on the result of measurements (i.e., $\ket{\uparrow}$ or $\ket{\downarrow}$), and the initial entangled states over elementary links (i.e., $\ket{\psi^{\pm}}$ given in Eq.\ref{eq:Bellstate}), the entangled state between the outer nodes will be projected onto a Bell state.

In the following, we analyze two different approaches to achieve the required interaction to perform a CNOT gate between ions. Performing a deterministic gate using the virtual exchange of photons is discussed in Sec \ref{sssec:virtual}. We also discuss how monitoring the cavity emission can improve the fidelity of this scheme.  In Sec. \ref{sssec:electric} we explain another scheme to perform a deterministic gate using the electric dipole-dipole interaction. 

\subsubsection{Virtual Photon Exchange}\label{sssec:virtual}
Since both Er ions of a single node are coupled to the same cavity, the interaction between these two ions can be mediated by the exchange of virtual cavity photons \cite{gate, blais}. Using this method, it is possible to perform a controlled phase-flip (CZ) gate between Er ions. A CZ gate combined with two Hadamard gates can then be used to perform a CNOT gate; $H_{Er_{i+1}}\otimes CZ_{Er_i,Er_{i+1}} \otimes H_{Er_{i+1}}$. 

\begin{figure}
	\includegraphics[width=8cm]{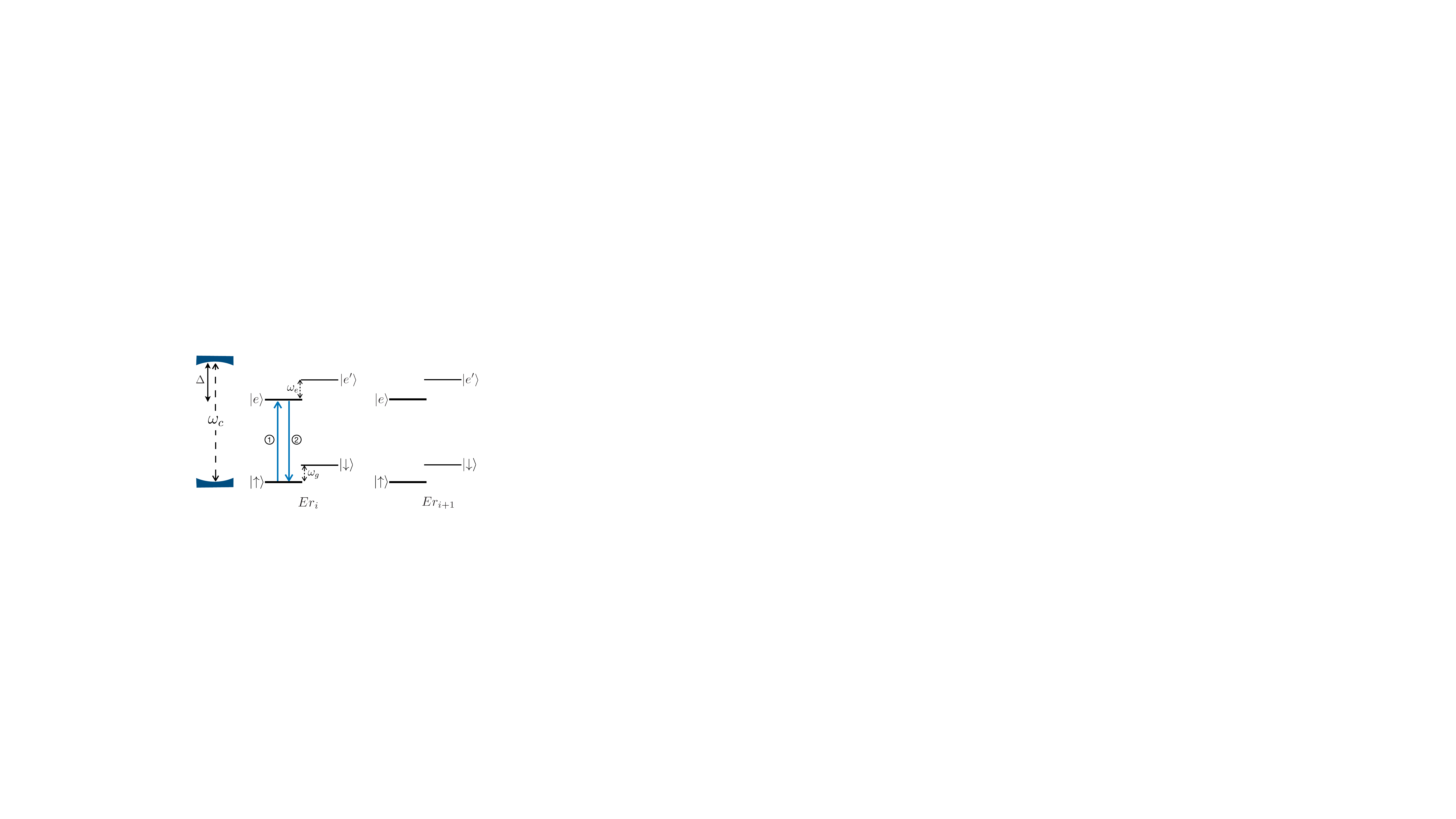}
	\caption{To perform a CZ gate using the virtual photon exchange, we bring the $\ket{\uparrow}$--$\ket{e}$ transition of the ions into resonance with each other while ions are dispersively coupled to the cavity. Using an optical $\pi$ pulse, we then excite one of the ions and let the exchange of virtual cavity photons perform a $\pi$ phase shift on the state.}\label{fig:lukin}
\end{figure}

To perform the CZ gate, the $\ket{\uparrow}$--$\ket{e}$ transitions of the ions are brought into resonance while both are dispersively coupled to a cavity mode (with the cavity detuning $\Delta$). Then, we excite the first ion using an optical $\pi$ pulse resonant with the $\ket{\uparrow}$--$\ket{e}$ transition, as shown in Fig.\ref{fig:lukin}. If the joint state of the ions was $\ket{\uparrow\uparrow}$, then after exciting the ion, the virtual exchange of a cavity photon between degenerate states $\ket{\uparrow e}$ and $\ket{e \uparrow}$ adiabatically performs a $\pi$ phase shift on the state.
Finally, another optical $\pi$ pulse brings the excited qubit back to the ground state after a delay time. 

So long as the splitting between states  $\ket{e\downarrow}$ and $\ket{\uparrow e'}$ (which is $\delta_{eg}=(\omega_e-\omega_g)$) is large enough and the system has negligible spin-flip transitions coupled to the cavity, the other joint states of ions will not be affected by the pulses 1 and 2 \cite{gate}.
The unitary operator of this phase-flip gate can be written as $U_{CZ_{(Er_i,Er_{i+1})}}=-\ket{\uparrow\uparrow} \!\bra{\uparrow\uparrow}+\ket{\uparrow\downarrow} \!\bra{\uparrow\downarrow}+\ket{\downarrow\uparrow} \!\bra{\downarrow\uparrow}+\ket{\downarrow\downarrow} \!\bra{\downarrow\downarrow}$.

After performing the CNOT gate, to complete the swapping process we measure $Er_i$ in the X basis and $Er_{i+1}$ in the Z basis. 

The two main processes limiting the fidelity of the CZ gate are cavity mode dissipation and spontaneous emission. If the cavity detuning is too small, the Purcell enhancement will cause the ions to decay into the cavity mode before the completion of the phase flip. On the other hand, if the detuning is too large, the dissipative interaction will be too slow to complete the phase flip before spontaneous emission occurs. The former limitation can be relaxed if the cavity emission is efficiently collected and monitored during the gate. Doing so allows for the rejection of gate attempts where cavity emission occurred, thus improving fidelity at the cost of some efficiency. Adding such a post-selection scheme also allows for the scheme to be performed with a smaller cavity detuning, which in turn, decreases the gate time and makes the scheme more robust against other decoherence processes.
 
\subsubsection{Electric dipole-dipole interaction} \label{sssec:electric}

Optically exciting an Er ion changes its permanent electric dipole moment. As a result, the electric field environment around the ion will change.
This change in the local electric field can impact 
other nearby ions by shifting their optical transition frequencies by \cite{altner}:
	\begin{equation}
	\label{eq:shift}
	\begin{aligned}
\Delta\nu=&\frac{\Delta\mu_{{Er}_i}\Delta\mu_{Er_{i+1}}}{4\pi\epsilon\epsilon_{0}hr^{3}}\\&\left(\left(\hat{\mu}_{Er_i}\!\cdot\!\hat{\mu}_{Er_{i+1}}\right)-3\left(\hat{\mu}_{Er_i}\!\cdot\!\hat{r}\right)\left(\hat{\mu}_{Er_{i+1}}\!\cdot\!\hat{r}\right)\right),
\end{aligned}
\end{equation}
where $Er_i$ is the excited ion, $Er_{1+i}$ is its nearby ion, $\Delta\mu$ is the change of the permanent electric dipole moment, $r$ is the distance between ions, $\epsilon_{0}$ is vacuum permittivity, $h$ is the Planck constant, and $\epsilon$ is the dielectric constant.
Using this modification in the transition frequency, one can perform a deterministic CNOT gate between nearby qubits.
\begin{figure}
	\includegraphics[width=7cm]{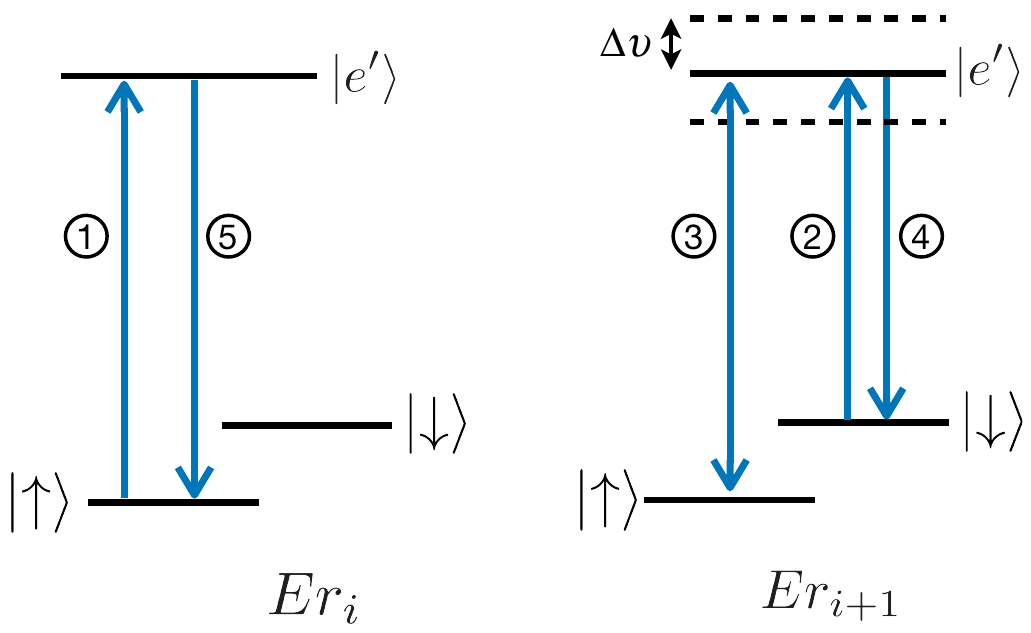}
	\caption{Pulse sequence to perform a CNOT gate between close-lying $Er_{i}$ (control) and $Er_{1+i}$ (target) ions. When $Er_{i}$ is in the state $\ket{\uparrow}$, due to the electric dipole-dipole interaction pulses 2-4 have no effect of the target ion. Please note that all pulses shown here are optical $\pi$ pulses. The double arrow for pulse 3 illustrates that the pulse can either take the state $\ket{\uparrow}$ to $\ket{e}$ or $\ket{e}$ to $\ket{\uparrow}$. }\label{fig:electric-dipole}
\end{figure}
For both ions, we consider that the transition $\ket{\uparrow}$--$\ket{e'}$ is detuned from the cavity.
First, we apply a short optical $\pi$ pulse resonant with the $\ket{\uparrow}$--$\ket{e'}$ transition of the control ion (e.g.,  Er$_i$), as shown in Fig. \ref{fig:electric-dipole}.
Then, using pulses 2--4, we swap population in the target ion.
Finally, pulse 5 brings the control ion back to its ground state.

This process can be interrupted by the electric dipole-dipole interaction if i) the control ion is in the state $\ket{\uparrow}$ and ii) the ions are sufficiently close to each other. In this case, if the shift in the transition frequency $\Delta\nu$ of the target ion is large compared to the transition linewidth, pulses 2--4 do not affect the system \cite{electric-dipole, electric-dipole2}.

The overall result of this interaction is that the state of the target qubit is flipped if the control qubit is in the state $\ket{\downarrow}$.
After performing the CNOT gate, we need to also measure the control (target) ion in the X (Z) basis.

\section{Fidelity and efficiency}\label{ssec:fidelity}
Here we estimate the fidelity and efficiency for each step as well as the end-to-end fidelity of the protocol. We also show numerically how the fidelity of the virtual photon exchange swapping gate can be improved by monitoring the cavity emission. In this section, fidelity is computed as $F=\bra{\psi}\hat{\rho}\ket{\psi}$, where $\hat{\rho}$ is the imperfect final state and $\ket{\psi}$ is the expected pure state.

\subsection{Entanglement generation} \label{BK-fidelity}
When spin decoherence is negligible on the time scale of the optical dynamics, and the system operates in the bad-cavity regime, the fidelity of the Barrett-Kok entanglement generation scheme between two ions identical other than for emission wavelength is given by \cite{conditional}:
\begin{equation}
\label{barrett-kok}
F_\text{entangle}= \frac{1}{2}\left(1 + \frac{\gamma^{\prime2}}{\Gamma^{\prime^2}+\Delta_w^2}\right),
\end{equation}
where $\gamma^\prime=\gamma_{r}F_p+\gamma$ is the Purcell-enhanced optical decay rate of the ion in the presence of the cavity, $\gamma=\gamma_r+\gamma_{nr}$ is the bare ion decay rate, $\gamma_r$ ($\gamma_{nr}$) is the radiative (non-radiative) component of the decay rate, $\Gamma^\prime=\gamma^\prime+2\gamma^\star$ is the FWHM of the Purcell-enhanced zero-phonon line (ZPL), and $\Delta_w$ is the difference between the optical transition frequencies of the ions. We define $\gamma^\star$ as the optical pure dephasing rate and $F_p=R/\gamma_r$
as the Purcell factor, where $R=4g^2/(\kappa+\Gamma)$ is the effective transfer rate of population between the ion and cavity \cite{grange2015cavity}, $\Gamma=\gamma+2\gamma^\star$ is the FWHM of the ZPL before enhancement, $g$ is the cavity-ion coupling rate, and $\kappa$ is the cavity decay rate. In the regime where $\kappa \gg \Gamma$, the Purcell factor can be written as $F_p=4g^2/(\gamma_r \kappa)$.

The entanglement generation fidelity is related to the mean wavepacket overlap $M^\prime$ of Purcell-enhanced photons from each ion $M^\prime = \Gamma^\prime\gamma^\prime/(\Gamma^{\prime^2}+\Delta_w^2)$ \cite{conditional} by $F_\text{entangle}=(1+M^\prime I^\prime)/2$ where $I^\prime=\gamma^\prime/\Gamma^\prime$ is the indistinguishability of the Purcell-enhanced photons from one ion. Note that the fidelity is less than or equal to $(1+M^\prime)/2$, which would be the expected fidelity when accounting for interference visibility only \cite{Hanson}. This is because the optical pure dephasing of the emitter degrades both the temporal coherence of the emitted photons and the spin coherence of the ion state. Knowing this, the quantity $I^\prime$ in $F_\text{entangle}$ actually accounts for the degredation of the ion spin coherence while $M^\prime\leq I^\prime$ quantifies the reduced interference visibility of photons from separate ions. In general, to have high interference visibility, spectral diffusion of the optical transition needs to be controlled. In our system, however, we expect the spectral diffusion to be negligible (see Sec.\ref{E-generation} for more information). 

\begin{figure}
    \centering
    \includegraphics[scale=0.6]{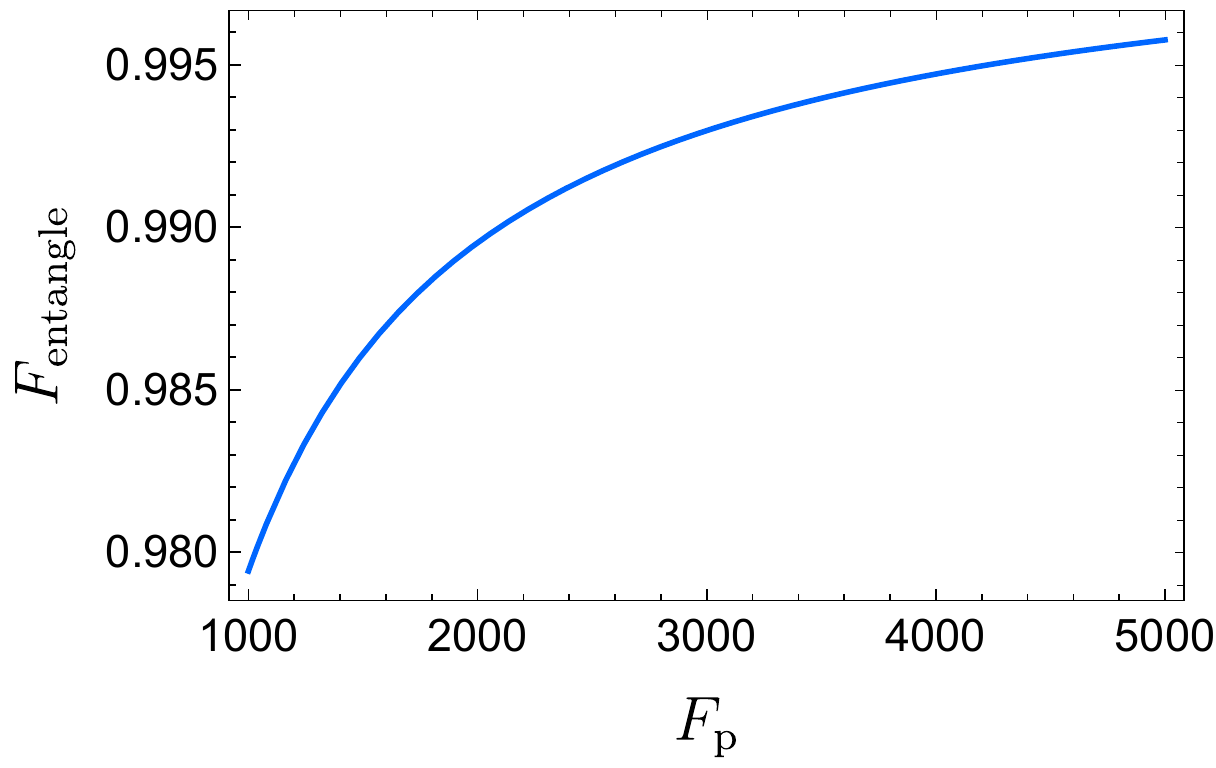}
    \caption{Fidelity of the entanglement generation scheme with respect to the Purcell factor. Here we assume $\Delta_w=0$.}
    \label{fig:BK}
\end{figure}

The presence of the cavity helps improve the single-photon indistinguishability as $I^\prime=I(1+\zeta F_p)/(1+I \zeta F_p)$ where $I=\gamma/\Gamma$ is the single-photon indistinguishability in the absence of the cavity and $\zeta=\gamma_r/\gamma$. This in turn improves the mean wavepacket overlap and consequently improves the entanglement generation fidelity (see Fig.~ \ref{fig:BK}).

We estimate $\gamma^\star$ using the relation $\gamma^\star=1/T_2-\gamma/2=2\pi\times32$ Hz,
where $T_2=4$ ms is the optical coherence time (for B=7 T) \cite{opticalT2}, and $\gamma=2\pi\times 14$ Hz \cite{opticalT1}.
Considering $\gamma_r=2\pi\times3$ Hz \cite{mcauslan2009strong}, $\gamma_{nr}=2\pi\times11$ Hz and $\Delta_\omega=0$ the entanglement generation fidelity would be $F_\text{entangle}=$0.996 for $F_p=5000$, as shown in Fig.~\ref{fig:BK}.

The entanglement
generation efficiency is given by $p_\text{en} = \eta^2/2$, where
$\eta=p\eta_t \eta_d$, $p=\eta_c F_p \gamma_r/\gamma^\prime$ is the success probability
of single-photon emission into a collection fibre mode (see Sec.~\ref{measurement}), $\eta_c$ is the collection efficiency, $\eta_{t}=e^{-\frac{L_{0}}{2L_{att}}}$ is the transmission efficiency in the fibre, $L_{att}\approx22$ km is the attenuation
length (corresponding to a loss of 0.2 dB$/$km), and $\eta_d$ is the detection efficiency. 

Before the first entanglement generation attempt and then again after every unsuccessful attempt, the ions must be initialized in the ground state $\ket{\uparrow}$. For an ion in a cavity with a large Purcell enhancement, the initialization fidelity after applying a single pulse exciting all ground states other than $\ket{\uparrow}$ to $\ket{e}$ is given by $F_\text{init}\geq(\gamma_rF_p+\beta\gamma)(1-e^{-T_\text{init}\gamma^\prime})/\gamma^\prime$ where $T_\text{init}$ is the initialization time needed for the excited state to decay and $\beta \geq 0.9$ is the branching ratio without a cavity \cite{hastings}. For $F_p=5000$ and $T_\text{init}=8/\gamma^\prime=85\mu$s the fidelity can be as high as $F_\text{init}\geq 0.9996$. This small infidelity is negligible compared to $F_\text{entangle}$, however $T_\text{init}$ can slightly reduce the repeater rate for small distances  (see Sec. \ref{ssec:rate} for more information). For this fidelity estimation, we ignore the ground state $T_1$ thermalization time, which is on the order of seconds \cite{sellars} and is negligible compared to the Purcell-enhanced lifetime.

\subsection{Virtual photon exchange}
\label{exchangegatesection}

While performing the CNOT gate,
there is always some infidelity due to the Hadamard gates
which do not depend on the scheme. Here we assume 
the fidelity of the CNOT gate will be dominated by the phase gate step. In the absence of excess dephasing and in the bad-cavity regime where $\kappa > g$, the cooperativity-limited gate fidelity is $F_\text{max}=1-2\pi/\sqrt{C}$ \cite{gate} where $C=4g^2/(\kappa\gamma) \gg 1$. This limit is reached when the ions are detuned from the cavity by $\Delta=\kappa\sqrt{C}/2$, which implies that the gate time $T_\text{gate}=\pi\Delta/g^2$ becomes $T_0=2\pi/(\gamma\sqrt{C})$ under the optimal conditions. Note that in this work we define fidelity to be consistent with Ref. \cite{conditional} and so it is the square of the fidelity as defined in Ref. \cite{gate}. Hence some expressions in this section differ from that of Ref. \cite{gate} accordingly.

A slight detuning $\Delta_w$ between ions within the cavity or incidental cavity coupling of another transition detuned by $\delta_{eg}$ can both cause imperfections in the phase evolution of the CZ gate. To lowest order for the infidelity contributions due to finite $C$, $\Delta_w$, and $\delta_{eg}$ the gate fidelity maximum is given by \cite{gate}
\begin{equation}
   \label{Lukin}
    F_\text{max} = 1 - \frac{2\pi}{\sqrt{C}}-\frac{6\pi^2}{32}\!\!\left[\left(\!\frac{T_0\Delta_w}{2\pi}\!\right)^2\!\!\!+\left(\!\frac{2\pi}{T_0\delta_{eg}}\!\right)^2\right].
\end{equation}
These imperfections reduce the maximum achievable fidelity but they do not change the ideal ion-cavity detuning. On the other hand, some excess dephasing can both reduce the maximum fidelity and favor shorter gate times, which in turn reduces the optimal ion-cavity detuning.

\begin{figure}
    \centering
     \hspace{-50mm}\textbf{a.}\\
     \hspace{-4mm}\includegraphics[scale=0.75]{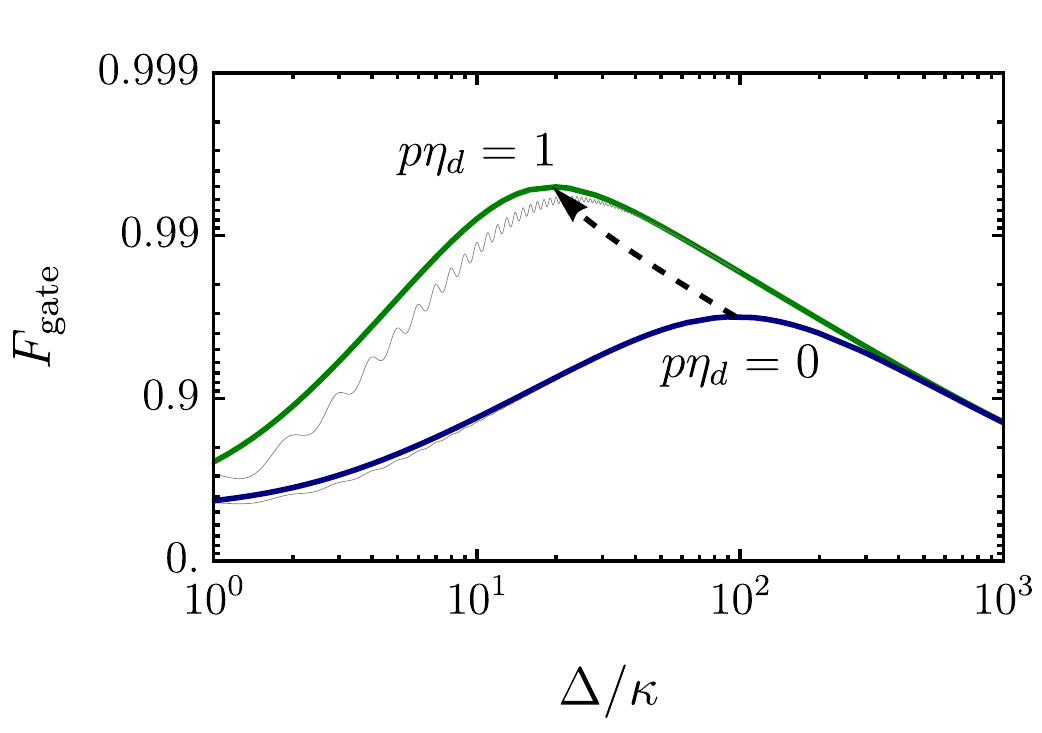}\\
     \hspace{-50mm}\textbf{b.}\\
    \hspace{-5.mm}\includegraphics[scale=0.75]{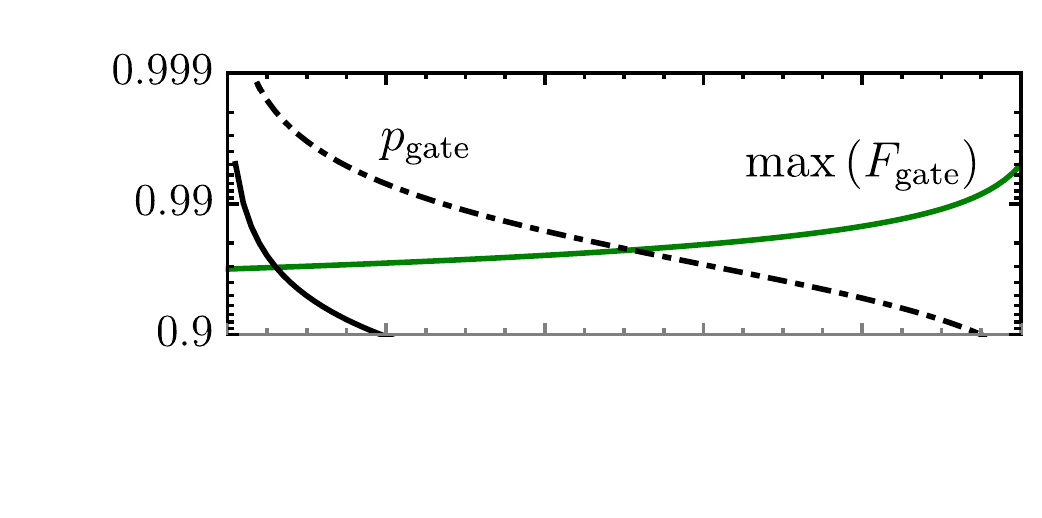}\\\vspace{-19.6mm}
    \includegraphics[scale=0.724]{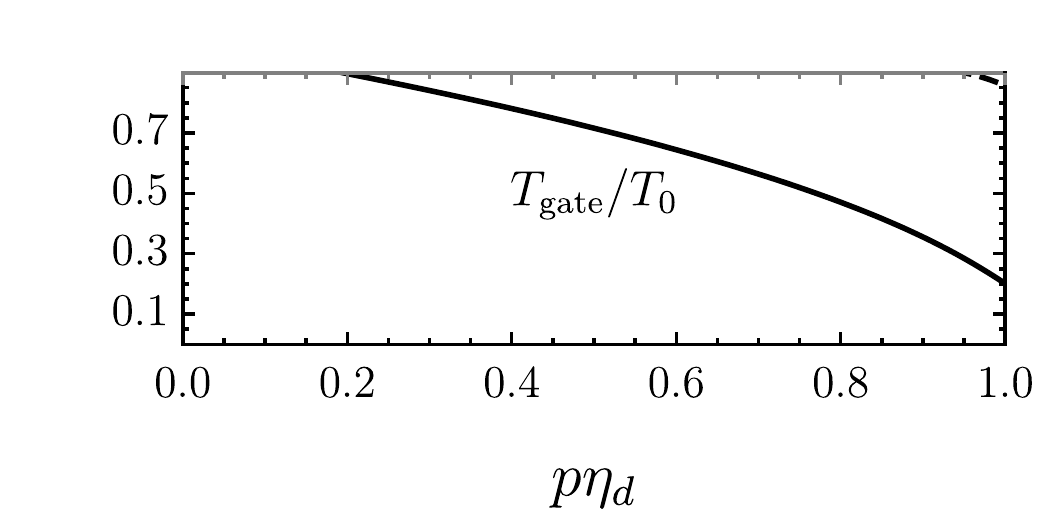}
    \caption{Cavity-mediated virtual photon exchange controlled phase-flip gate. \textbf{a.} Phase-flip gate fidelity $F_\text{gate}$ as a function of cavity detuning $\Delta$ for no post-selection (blue curve, $p\eta_d=0$) and perfect post-selection (green curve, $p\eta_d=1$) of gates where no emission was observed from the cavity. The black dashed arrow shows the path that the fidelity peak follows when increasing the cavity monitoring efficiency $p\eta_d$ from 0 to 1 in the direction of the arrow. The thin gray lines show the case where $\delta_{eg}=\kappa/50$ as opposed to $\delta_{eg}\gtrsim \kappa$ of the colored solid lines. \textbf{b.} The maximum fidelity for a given monitoring efficiency $p\eta_d$ corresponding to the dashed arrow in panel a. plotted alongside the corresponding gate efficiency $p_\text{gate}$ and gate time $T_\text{gate}/T_0$ where $T_0$ is the optimal gate time for the case of no post-selection ($p\eta_d=0$). The cavity mediating the interaction is in the bad cavity regime with $g/\kappa=10^{-1}$ and a cooperativity of $C=9\times10^4$. For this simulation we assumed an optical pure dephasing rate of $\gamma^\star = 2.3 \gamma$ for both ions.}
    \label{fig:lukinpostselect}
\end{figure}

Although the methods used in Ref. \cite{gate} cannot directly analytically account for the excess optical pure dephasing of Er ions, in the regime where the gate time $T_\text{gate}$ is small compared to $1/\gamma^\star$, the resulting infidelity is proportional to $T_\text{gate}\gamma^\star$. Using the solution for the detuning-dependent fidelity of Ref. \cite{gate}, we find that the gate fidelity for when $T_\text{gate}\Delta_w\ll2\pi$ and $T_\text{gate}\delta_{eg}\gg2\pi$ that includes pure dephasing is closely approximated by
\begin{equation}
\label{fidelitygate}
    F_\text{gate} =\frac{1}{4}(e^{-2\pi\Delta/C\kappa-\pi\kappa/2\Delta}+1)^2 - 0.29\gamma^\star T_\text{gate},
\end{equation}
where $C=4g^2/(\kappa\gamma)$ is the cavity cooperativity when neglecting pure dephasing and $T_\text{gate}$ can be written in terms of $C$ as $T_\text{gate}=4\pi\Delta/(C\kappa\gamma)$. The coefficient 0.29 is an estimate obtained by comparing the analytic approximation to the numeric solution from simulating the master equation far in the bad-cavity regime. 

Alternatively, optical pure dephasing can be seen to effectively reduce the cavity cooperativity. However, because pure dephasing does not affect the gate fidelity in the same way as spontaneous emission, replacing $C$ by the usual definition for the reduced cavity cooperativity $4g^2/(\kappa\Gamma)$ is not accurate. Instead, we find that replacing $C$ by the quantity $C^\star=C\gamma/(\gamma+0.61\gamma^\star)$ provides an accurate analytic approximation, where again the coefficient 0.61 is estimated by comparison to the full numeric solution. The fact that 0.61 is smaller than 2 suggests that pure dephasing has much less of a degrading effect than would be naively expected. The maximum fidelity when accounting for pure dephasing then becomes $F_\text{max} = 1 - 2\pi/\sqrt{C^\star}$ and is achieved at the detuning $\Delta=\kappa\sqrt{C^\star}/2$ which implies an optimal gate time of $T_0 = \pi\kappa\sqrt{C^\star}/(2g^2)=2\pi\sqrt{C^\star}/(C\gamma)$.

For a given cavity cooperativity, the maximum fidelity for the virtual photon exchange gate can be increased if successful attempts are post-selected when no cavity emission is observed during the interaction. To estimate the amount of improvement, we numerically simulated the state of the system given that a detector monitoring the cavity mode emission did not measure a photon. 

We consider the Hamiltonian in \cite{gate}: $\hat{H} = \hat{H}_A + \hat{H}_B+\hat{H}_C + \hat{H}_I$ where $\hat{H}_k$ is the $k^\text{th}$ ion Hamiltonian, $\hat{H}_C$ is the cavity mode Hamiltonian and $\hat{H}_I$ is the cavity-ion interaction. The four-level ion Hamiltonian is
\begin{equation}
    \hat{H}_k = \omega_k\hat{\sigma}_{\uparrow_k}^\dagger\hat{\sigma}_{\uparrow_k}+(\omega_k+\omega_e)\hat{\sigma}_{\downarrow_k}^\dagger\hat{\sigma}_{\downarrow_k}+\omega_g\hat{\sigma}_{\uparrow\downarrow_k}^\dagger\hat{\sigma}_{\uparrow\downarrow_k},
\end{equation}
where $\omega_k$ is the frequency separation between $\ket{\uparrow}_k$ and $\ket{e}_k$, $\omega_e$ is the separation between $\ket{e}_k$ and $\ket{e^\prime}_k$, and $\omega_g$ is the separation between $\ket{\uparrow}_k$ and $\ket{\downarrow}_k$. Also, $\hat{\sigma}_{\downarrow_k}\ket{e^\prime}_k=\ket{\downarrow}_k$, $\hat{\sigma}_{\uparrow_k}\ket{e}=\ket{\uparrow}_k$, and $\hat{\sigma}_{\uparrow\downarrow_k}\ket{\downarrow}_k=\ket{\uparrow}_k$ (see figure \ref{fig:lukin}). The cavity homogeneous evolution is $\hat{H}_C = \omega_c\hat{a}^\dagger\hat{a}$ for cavity frequency $\omega_c$, cavity photon creation (annihilation) operator $\hat{a}^\dagger$ ($\hat{a}$), and the interaction term is
\begin{equation}
    \hat{H}_I = \sum_{j\in{\uparrow,\downarrow}}\sum_{k\in{A,B}}g_{j_k}\hat{\sigma}_{j_k}^\dagger\hat{a} + \text{h.c.},
\end{equation}
where $g_{\downarrow_k}$ is the cavity coupling rate of the $\ket{\downarrow}$--$\ket{e^\prime}$ transition to the cavity mode and $g_{\uparrow_k}$ is the cavity coupling rate of the $\ket{\uparrow}$--$\ket{e}$ transition to the cavity mode. In addition to the spontaneous emission rate $\gamma_k$ and cavity linewidth $\kappa$, we explicitly include an optical pure dephasing rate $\gamma^\star$ in the total Lindblad master equation given by
\begin{equation}
\begin{aligned}
    \dot{\rho} &= -i[\hat{H},\hat{\rho}] +\kappa\mathcal{D}(\hat{a})\hat{\rho}+ \sum_{k,j}\gamma_{j_k}\mathcal{D}(\hat{\sigma}_{j_k})\hat{\rho}\\
    &+2\gamma^\star_k\mathcal{D}(\hat{\sigma}_{\uparrow_k}^\dagger\hat{\sigma}_{\uparrow_k}+\hat{\sigma}_{\downarrow_k}^\dagger\hat{\sigma}_{\downarrow_k})\hat{\rho},
\end{aligned}
\end{equation}
where $\mathcal{D}(\hat{A})\hat{\rho}=\hat{A}\hat{\rho}\hat{A}^\dagger -\{\hat{A}^\dagger\hat{A},\hat{\rho}\}/2$. This master equation defines the superoperator $\mathcal{L}$ where $\dot{\rho} = \mathcal{L}\hat{\rho}$. 

Using the method of conditional evolution \cite{carmichael, zoller, conditional} the unnormalized conditional state $\hat{\rho}_0(t)$ at time $t$ given that no emission was observed from the cavity since time $t_0$ is
\begin{equation}
\label{conditionalstate}
    \hat{\rho}_0(t)=e^{(t-t_0)(\mathcal{L}-p\eta_d\kappa\mathcal{S})}\hat{\rho}(t_0),
\end{equation}
where $\mathcal{S}\hat{\rho}=\hat{a}\hat{\rho}\hat{a}^\dagger$ is the cavity photon collapse superoperator, $p$ is the probability of receiving a photon emitted by the cavity and $\eta_d$ is the detector efficiency. Then the probability that no photon is emitted from the cavity during the gate duration $t-t_0=T_\text{gate} = \pi\Delta/g^2$ is
\begin{equation}
   p_\text{gate}(T_\text{gate}) = \text{Tr}(\hat{\rho}_0(T_\text{gate})),
\end{equation}
where we assume that $g=g_{j_k}$ is the same for all transitions. In this case the final state after a successful gate is
\begin{equation}
    \hat{\rho}_\text{gate}=\frac{1}{p_\text{gate}}\hat{\rho}_0(T_\text{gate}).
\end{equation}

Here assuming $\kappa=2\pi \times 16$ MHz, for a 
quality factor of $1.2\times10^7$ and $C\simeq 9 \times10^4$, a cavity with a length of $\sim5\,\mu$m is required (for more information see Sec.\ref{E-generation}). In the bad-cavity regime where $g/\kappa=10^{-1}$, perfect monitoring efficiency $p\eta_d=1$ improves the maximum gate fidelity from $0.968$ to $0.995$ while also decreasing the optimal detuning from about 100$\kappa$ to 20$\kappa$, corresponding to a decrease in optimal gate time from $T_\text{gate}=160 \mu$s to $T_\text{gate}=32 \mu$s (see figure \ref{fig:lukinpostselect}). These improvements come at the cost of the scheme becoming non-deterministic with an efficiency of $0.86$. 

\subsection{Electric dipole-dipole interaction}
The achievable fidelity for this CNOT gate is \cite{Er-Eu}:
  \begin{equation}
    \label{dipole}
   		F_\text{gate} \simeq 1-\frac{T_\text{gate}}{80}\left(42\gamma+25\gamma^\star+25\chi\right)-\frac{43\pi^2}{128}\left(\frac{\delta\nu}{\Delta\nu}\right)^2,
    \end{equation}
where $T_\text{gate}=5\pi/\Omega=5\pi\sqrt{3}/\Delta\nu$ is the gate time, $\Omega$ is the Rabi frequency for optical transitions (here we assumed $\Omega_{\uparrow}=\Omega_{\downarrow}=\Omega$), $\Delta\nu$ is the shift in the transition frequency, $\delta\nu$ is a potential mischaracterization from the true value of $\Delta\nu$, and $\chi$ is the spin decoherence rate of the ion.
Eq.~\ref{dipole} is valid to first order in $\gamma,\gamma^\star,\chi\ll \Omega\propto\Delta\nu$ and second-order in $\delta\nu/\Delta\nu\ll1$.
\begin{figure}
    \centering
    \includegraphics[scale=0.6]{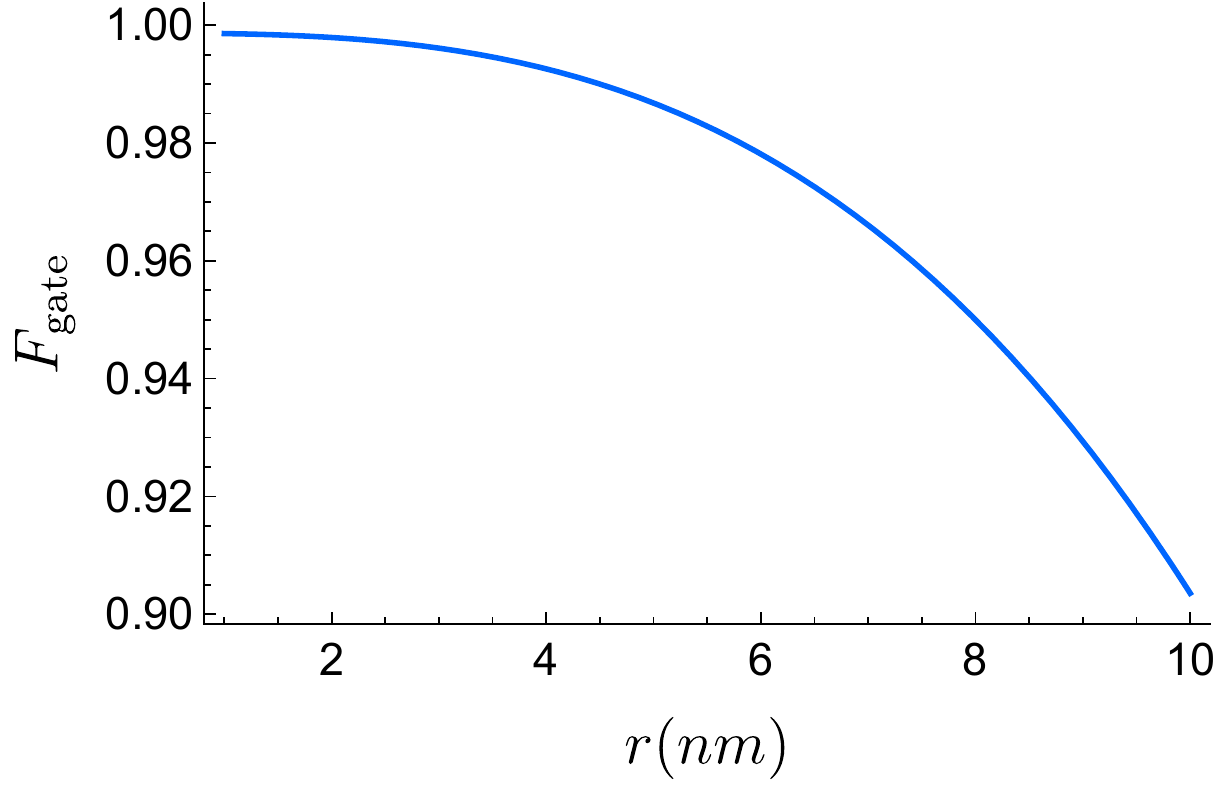}
    \caption{Fidelity of the electric dipole-dipole interaction gate as a
function of the distance between ions. Here we assume $\delta\nu/(\Delta\nu)=0.02$.}
    \label{fig:dipole-fidelity}
\end{figure}
Considering $\chi=2\pi\times0.12$ Hz (for B=7 T) \cite{sellars}, $\gamma=2\pi\times14$ Hz \cite{opticalT1}, $\gamma^\star=2\pi\times32$ Hz, $\Delta\nu=250$ KHz (corresponding to $r=5$ nm) and $\delta\nu/\Delta\nu=0.02$ the fidelity and gate time are $F_\text{gate}=0.987$ and $T_\text{gate}=108\,\mu$s, respectively. Fig.~\ref{fig:dipole-fidelity} shows the gate fidelity as a function of the separation between ions. 

\subsection{State readout}\label{readout}

For state readout, we assume that the $\ket{\uparrow}$--$\ket{e}$ transition is resonant with the cavity while $\ket{\downarrow}$--$\ket{e^\prime}$ is not. By exciting $\ket{\uparrow}$ with a sequence of $\pi$-pulses and monitoring emission as it decays back to $\ket{\uparrow}$, the intensity contrast can be used to distinguish $\ket{\uparrow}$ from $\ket{\downarrow}$ with high fidelity \cite{cycle, raha}.

Let B (D) denote the measurement result indicating $\uparrow$ ($\downarrow$). Then the fidelity is the conditional probability $P(\uparrow\!\!|B)$ ($P(\downarrow\!\!|D)$) of being in state $\uparrow$ ($\downarrow$) given the measurement outcome B (D). The total probability of success is $p_\text{readout}=P(B)+P(D)$ and we define the total fidelity as the weighted average of the conditional fidelity $F_\text{readout}=(P(B)P(\uparrow\!\!|B) + P(D)P(\downarrow\!\!|D))/p_\text{readout}$. If we assume that $P(\uparrow)=P(\downarrow)=1/2$ and events B and D are complementary resulting in a deterministic scheme where $p_\text{readout}=1$, then using Bayes' theorem: $P(i|j)=P(i)P(j|i)/P(j)$ where $i\in\{\uparrow,\downarrow\}$ and $j\in\{\text{B, D}\}$, we have $F_\text{readout} = (P(B|\!\uparrow)+P(D|\!\downarrow))/2$.

Suppose that the detector has a probability of $\xi\ll1$ to detect a photon when the ion is in state $\ket{\downarrow}$ during a single cycle; for example, due to noise from excitation or dark counts. On the other hand, there is a chance $p\eta_d\gg \xi$ that a single cycle of $\ket{\uparrow}$--$\ket{e}$ results in a single-photon detection. A simple readout scheme is then to excite the ion a fixed number of $N$ times and define event B to be the detection of one or more photons and event D as the complimentary (no photons). The state $\ket{\downarrow}$ results in D if no detection occurs and so $P(D|\!\downarrow)=(1-\xi)^N\simeq 1-N\xi$. On the other hand, the probability to see at least one photon after $N$ pulses is given by $P(B|\!\uparrow)=1-(1-p\eta_d)^N$. Hence $F_\text{readout} = 1 - N\xi/2-(1-p\eta_d)^N/2$.

\begin{figure}
    \centering
    \includegraphics[scale=0.63]{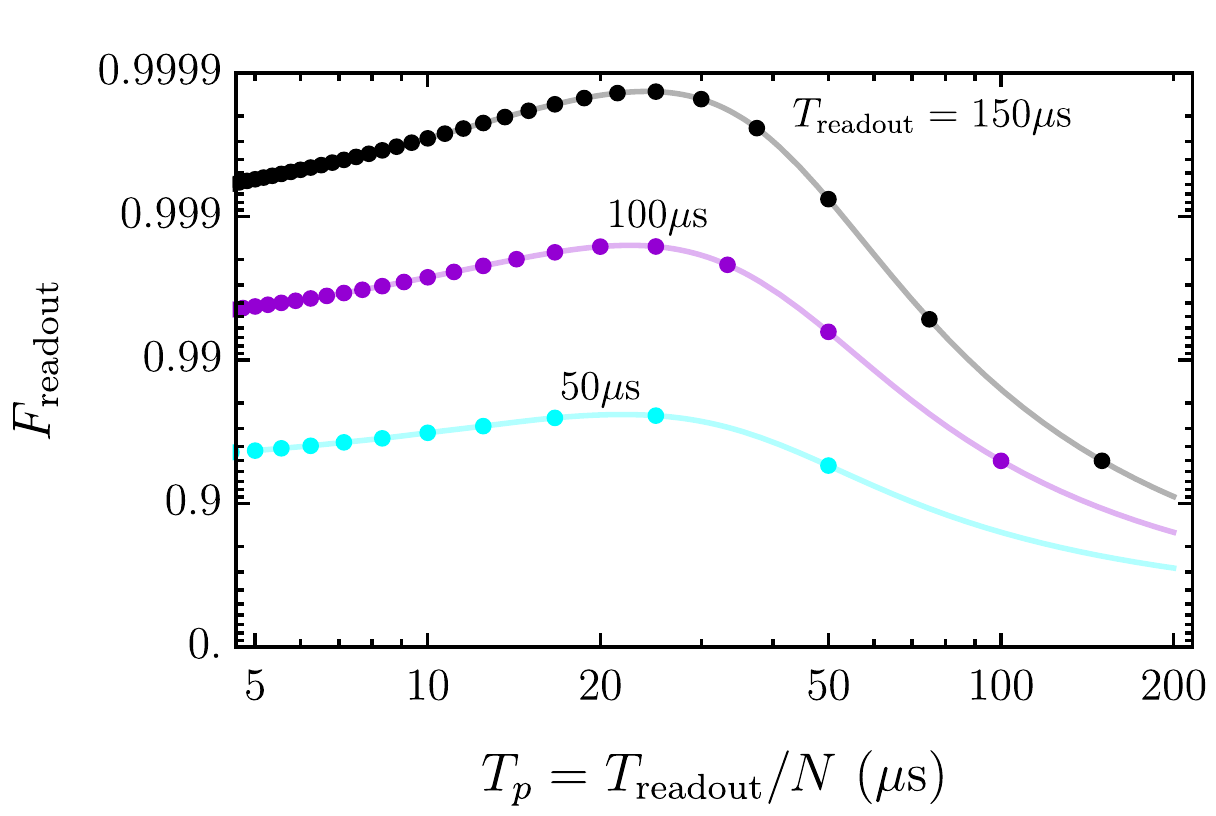}
    \caption{Readout fidelity $F_\text{readout}$ as a function of pulse separation $T_p$ for $p\eta_d=0.9$, $\xi=10^{-5}$, and a Purcell factor of $F_p=5000$ corresponding to $1/\gamma^\prime=10.6\mu$s. The points show the valid discrete values associated with an integer number of total pulses $N$. The curves connecting the points are given by the continuous extension of $F_\text{readout}$ to help guide the eye.}
    \label{figreadout}
\end{figure}

For a fixed number of $N$ pulses with a repetition period of $T_p$, the fixed readout time is $T_\text{readout}=NT_p$ when assuming that the measurement time after the last pulse is also $T_p$. To optimize this readout time, it is possible to increase the repetition rate. However, doing so risks coherently de-exciting the ion when $T_p\lesssim 1/\gamma^\prime$, effectively reducing the probability of emitting a photon. Considering that the probability for decay between pulses separated by $T_p$ is $1-e^{-T_p\gamma^\prime}$ we find that the probability for emission between the $k$th and $(k+1)$th pulse is $\eta_p(k)=(1-(-1)^k e^{-kT_p\gamma^\prime})\tanh(T_p\gamma^\prime/2)$. Then the fidelity becomes $F_\text{readout} \simeq 1 - N\xi/2-(1/2)\prod_{k=1}^N[1-p\eta_d\eta_p(k)]$. Using a cavity with $F_p=5000$, the ion lifetime is $1/\gamma^\prime\simeq 10.6\mu$s. Then for a readout time of $T_\text{readout}=150\mu$s using $N=7$ pulses with $T_p=21.4\mu$s, the fidelity can be as high as $F_\text{readout}=0.9998$ for $p\eta_d=0.9$ and $\xi=10^{-5}$ (see Fig. \ref{figreadout}).


\subsection{End-to-end repeater fidelity}\label{overal-fidelity}
To estimate the fidelity of the final entangled state over the entire distance $L$, we multiply the fidelities of all the
individual steps for a repeater protocol as following

\begin{equation}
\label{end-to-end}
    F_\text{end-to-end}=\left(F_\text{init}\right)^{2m} \times\left(F_\text{entangle}\right)^m \times\left(F_\text{swap}\right)^{m-1},
\end{equation}
where $m$ is the number of elementary links of length $L_0$. Note that Eq.\ref{end-to-end} is an estimation of the end-to-end fidelity that is only accurate at high fidelity regime. However, it still gives us a good approximation of the relative fidelity of the different cases. 
Here $F_\text{swap}$ includes two spin read-out measurements; therefore, $F_\text{swap}=F_\text{gate}\times (F_\text{readout})^2$ where $F_\text{gate}$ is the fidelity of performing the swapping gate for each scheme (i.e., Eqs. (\ref{Lukin}) and (\ref{dipole})). 
 The fidelity of the entanglement generation needs to be established over $m$ elementary links of length $L_0$.
 It has been shown that, even without the use of error correction protocols, the coherence time of $1$ s is more than enough to distribute entanglement over the distance of $L=1000$ km \cite{razavi}. 
Hence, we neglect the effect of the finite coherence time of the quantum memory due to the long hyperfine coherence time of the ${^{167}}$Er ion \cite{sellars}. 

\begin{figure}
    \centering
    \includegraphics[scale=0.6]{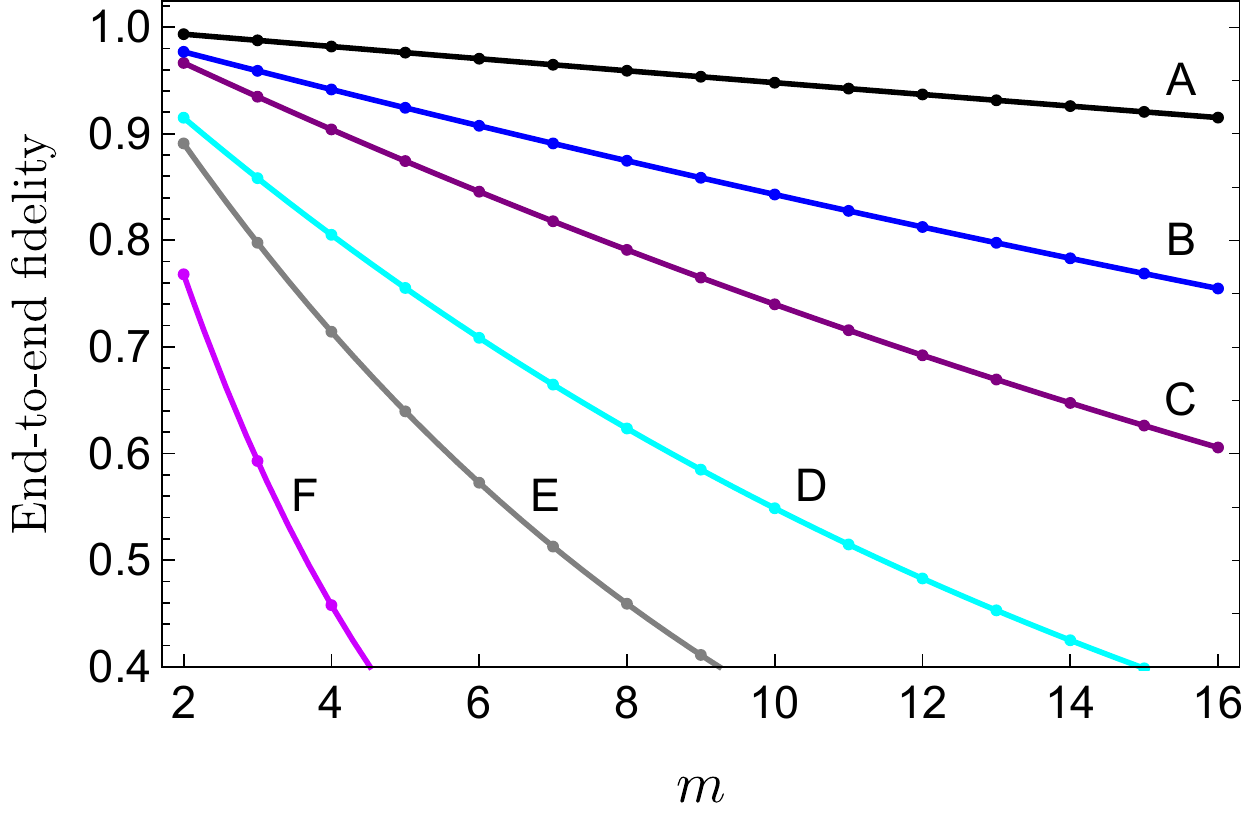}
    \caption{Estimation of the end-to-end fidelity of the repeater scheme with respect to the number of elementary links. 
    Shown are  the virtual photon exchange scheme post-selected on no cavity emission for $F_p=4.5\times10^5$ (A), and $F_p=5\times10^3$ (E), the virtual photon exchange scheme (without post-selection) for $F_p=4.5\times10^5$ (C), and $F_p=5\times10^3$ (F), the electric dipole-dipole interaction scheme for $F_p=5\times10^3$ (B), and also the scheme of Ref \cite{Er-Eu} for $F_p=5\times10^3$ (D). Here we assumed  $p\eta_d=0.9$, $N=7$, $T_p=2/\gamma^\prime$, $\xi=10^{-5}$, $\Delta_w=0$ and $T_\text{init}=8/\gamma^\prime$. For europium (Eu) ions in ref \cite{Er-Eu}, we also assumed $\gamma_{Eu}=2\pi\times 80$ Hz, $\gamma^\star_{Eu}=2\pi\times19$ Hz and $\chi_{Eu}=0$ \cite{mcauslan2009strong,6hours}.}
    \label{fig:overall-fidelity}
\end{figure}

Fig.~\ref{fig:overall-fidelity} shows the end-to-end fidelity estimation of the repeater protocol for different swapping schemes studied here as a function of the number of elementary links $m$ for $F_p=4.5\times10^5$ and $F_p=5\times10^3$. As shown, the end-to-end fidelity of virtual photon exchange scheme increases significantly by monitoring the cavity emission to post-select successful gates.
We have also shown the end-to-end fidelity of the scheme of Ref \cite{Er-Eu}. In this case the Eq.~(\ref{end-to-end}) changes to $F_\text{end-to-end}=\left(F_\text{init}\right)^{3m+1}\times\left(F_\text{entangle}\right)^m \times\left(F_\text{swap}\right)^{m-1}\times \left(F_\text{map}\right)^{m/2}$ where $F_\text{swap}=F_\text{map}=\left(F_\text{gate} \times F_\text{readout}\right)^2$ as each of these steps require performing two CNOT gates and two measurements (note that for the entanglement mapping, one needs to perform the gate and measurement at each end of the link) \cite{Er-Eu}. To further increase the fidelity of the repeater schemes, purification protocols could be used \cite{dur1999quantum}.

\section{Entanglement generation rate}\label{ssec:rate}
The average time to distribute entanglement over two elementary links of length $L_0$ is \cite{sangouard}
	\begin{equation}
		\braket{T}_{2L_0}=\left(\frac{3}{2}\right) \frac{L_0/c+T_\text{init}}{p_{en}p_s},
	\end{equation}
where $c=2\times10^8 \mathrm{\frac{m}{s}}$, $p_{en}$ is the success probability of entanglement generation over an elementary link (see Sec.\ref{BK-fidelity}), and $p_s=p_\text{gate}$ is the success probability of the entanglement swapping. 
The entanglement generation time over the entire distance $L=m L_0$, where $m$ is the number of links, is then given by \cite{NVroomT}
\begin{equation}
\label{time}
    \langle T \rangle_L=f(m)\frac{L_0/c+T_\text{init}} {\, p_{en} \,p_s^{m-1}},
\end{equation}
where $f(m)=0.64\log_2(m) + 0.83$ is a good approximation of the average number of attempts to successfully generate entanglement (this factor reduces to $3/2$ for the case of $m=2$).
\begin{figure}
    \centering
    \includegraphics[scale=0.6]{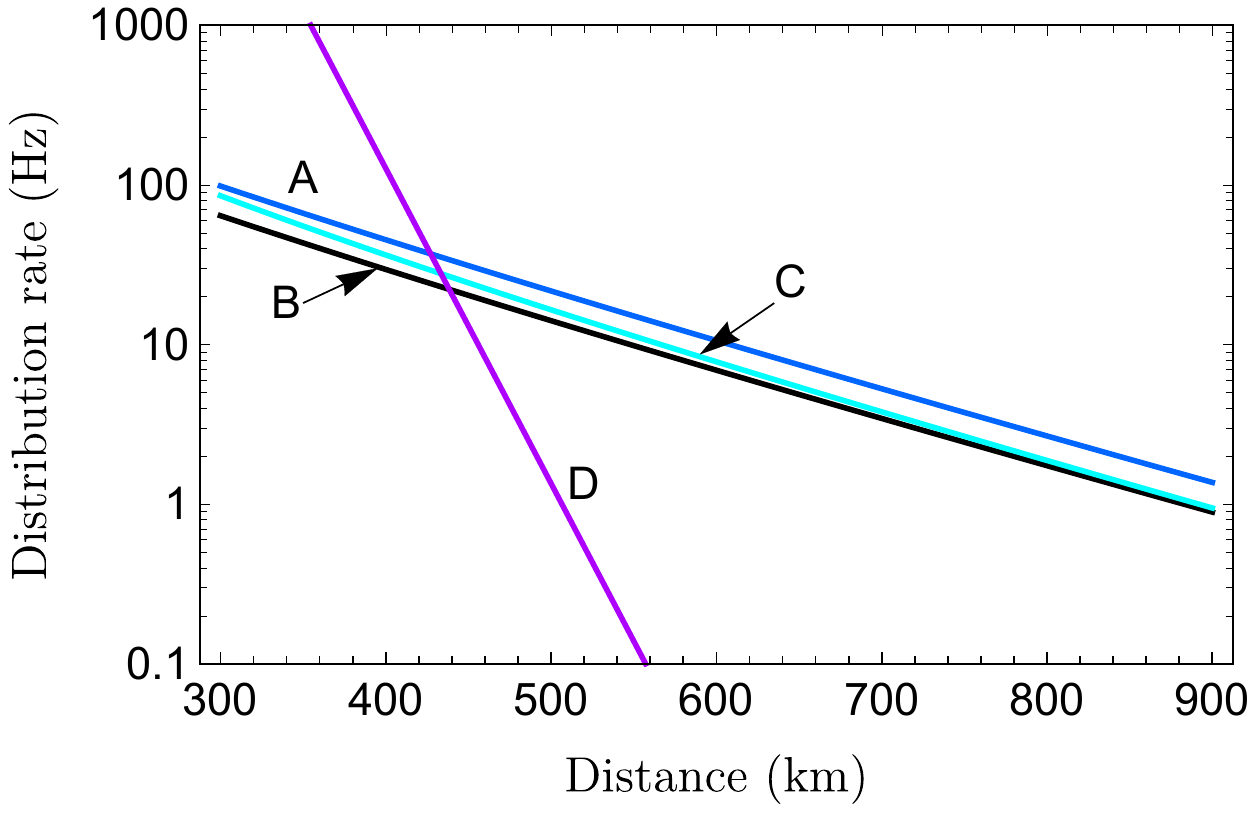}
    \caption{Comparison of the entanglement generation rate as a function of the distribution distance for single rare-earth ion-based repeater protocols. Our protocol for the deterministic entanglement swapping ($p_\text{gate}=1$) (A) is compared with the the protocol of the Ref.\cite{Er-Eu} (B). Also shown is our protocol for the probabilistic swapping gate ($p_\text{gate}=0.93$) (C), and finally the direct transmission in a fiber with a 1 GHz single-photon source (D). Here we assumed $m=8$ and $T_\text{init}=8/\gamma^\prime$. For lines A and B we set $F_p=5\times 10^3$ and for line C we assumed $F_p=4.5\times 10^5$.}
    \label{fig:rate}
\end{figure}

Here we assumed that the entanglement generation process could be performed on neighboring links at the same time. On the other hand, if entanglement generation should be established on neighboring links one by one (which is the case when spatial resolution of ions is not possible), then Eq.~(\ref{time}) changes to  
\begin{equation}
\label{time2}
    \langle T \rangle_L=2f(m/2)\frac{L_0/c+T_\text{init}} {\, p_{en} \,p_s^{m-1}},
\end{equation}

	
Note that here the swapping time (which includes the times required to perform the gate and read out the ions) is negligible compared to the time it takes to establish two neighboring links. As an example, even for the extreme case of $L=300$ km and $m=8$, the time needed to establish two neigboring links is $3.82$ ms, which is quite large compare to the gate and readout times discussed in Sec.\ref{ssec:fidelity}. This waiting time increases significantly by increasing the length of the elementary links. For instance, when $L=500$ km and $m=8$, the waiting time would be $19.83$ ms.

In Fig.\ref{fig:rate}, using Eq.(\ref{time}) we have plotted the entanglement generation rates of our proposed scheme as a function of distance for $m=8$ and compared the result with the rates achieved using the single Er-Eu scheme \cite{Er-Eu}. Line (A) shows the rate of our protocol for deterministic gates with $p_\text{gate}=1$ (i.e., virtual photon exchange without cavity monitoring or the electric dipole interaction), while (B) shows the rate for the protocol of Ref.\cite{Er-Eu}. Shown is also the repeater rates for the virtual photon exchange scheme post-selected on no cavity emission (C). Here we put $p_\text{gate}=0.93$ which corresponds to the  
$F_p=4.5\times 10^5$ and $p\eta_d=0.9$, as shown in Fig.\ref{fig:lukinpostselect}. 
Note that, in terms of the efficiency, a high Purcell factor is not required for the other schemes.
We have also plotted the rate expected using the direct transmission of photons with a 1 GHz photon rate (D) \cite{direct}. 

In the scheme of Ref. \cite{Er-Eu}, the process of measuring the communication ion ($^{168}$Er), after mapping its state to the memory ion ($^{151}$Eu), introduced an additional source of inefficiency to the system. In our proposed scheme, however, the single $^{167}$Er ions serve as both communication and memory ions; thus, the scaling with distance is better. By increasing the number of cycling transitions, the success probability of the measurement set improves, and the difference in rates between the two schemes becomes less drastic. However, even in this case, the implementation of the current proposed scheme is more experimentally feasible than the Er-Eu scheme because it does not require fabricating and identifying a close-lying pair of two species of ions.

\section{Implementation}
\label{ssec:implementation}
\subsection{Entanglement generation} \label{E-generation}
To perform entanglement generation between $Er_i$ and $Er_{i-1}$, for example, as illustrated in Fig\ref{fig:main}.a, we need to selectively optically address one ion at a time. Therefore, 
we either need to spatially address ions of the same cavity or put $Er_{i+1}$ and $Er_{i-2}$ ions out of resonance with the cavities they are placed in. One option to achieve the latter is through applying an electric field gradient to each cavity-ion system \cite{CRIB}. The Stark shift will then change the optical transition frequency of the ion out of resonance with the cavity.

Then, we need to prepare each ion in the ground state $\ket{\uparrow}$. Using frequency selection, pumping of $95\pm 3\%$ of the population into the $m_{I}=\frac{7}{2}$ hyperfine state has been demonstrated for an ensemble of Er ions \cite{sellars}. For an individual Er in the presence of a high-Purcell-factor cavity, a much higher percentage is expected.

The entanglement generation step also requires the excitation of the  $\ket{\uparrow}$--$\ket{e}$ transition for the ion which is resonant with the cavity. In order to avoid exciting both ground states to their respective excited states, the pulse spectral width should be much less than the difference between the ground and excited hyperfine level splittings. A cavity with a sufficiently small linewidth can also improve the branching ratio by enhancing one of the two transitions. For example, for the ground and excited states splitting difference of $\delta_{eg}\simeq 2\pi\times100$ MHz \cite{sellars}, a cavity with a linewidth of $\kappa$ (FWHM) centered on one transition can enhance that transition $\sim 1+4(\delta_{eg}/\kappa)^2$ times more than the transition detuned by $\delta_{eg}$. 

For rare-earth ions doped YSO photonic crystals cavities, quality factor of $27,000$ has been demonstrated \cite{zhong2016high}. Theoretical predictions, however, expect the quality factor as high as $10^5$, which could be improved even further by, for example, post-fabrication annealing or using materials with higher refractive indexes \cite{tian2009optical, zhong2016high}. Besides, quality factor exceeding $1.1\times 10^7$ has been measured in silicon photonic crystal nanocavities \cite{asano2017photonic}.

For Er ions doped into YSO crystal, B{\"o}ttger et al \cite{bottger2006optical} showed that the spectral diffusion decreases with increasing external magnetic field and decreasing temperature, and is undetectable at $B=3$ T and $T=1.6$ K even for a non-negligible Er concentration i.e., $0.0015\%$ (see Fig.2.c of the Ref \cite{bottger2006optical}). The spectral diffusion should be even lower under our conditions, where we only deal with individual Er ions at high fields and low temperatures. Hence, in this paper, we assume that the spectral diffusion is negligible compared to the $\Gamma^\prime$ (i.e., Purcell enhanced ZPL that dictates the entanglement generation fidelity), and $\Delta$ (i.e, ion-cavity detuning for swapping schemes). Note that, for the entanglement generation scheme, the Purcell effect further reduces the impact of spectral diffusion.  

\subsection{Entanglement swapping}\label{discussion}
In the following, we discuss pros and cons for each entanglement swapping scheme in more detail.

\textit{Virtual photon exchange}: Using this scheme, one can perform a deterministic gate between ions without the need of ions to be close-lying. During entanglement generation, the $\ket{\uparrow}$--$\ket{e}$ transition of an ion should be in resonance with the cavity. However, to perform the entanglement swapping using the virtual photon exchange scheme, the ions need to be dispersively coupled to the cavity. Therefore, one needs to detune the $\ket{\uparrow}$--$\ket{e}$ transitions away from resonance with the cavity before performing the entanglement swapping. An applied electric field amplitude $E$, could DC Stark shift the optical transition frequency of the ion by $\Delta=(\vec{\Delta\mu} \cdot \vec{E}) \alpha/\hbar$ where $\alpha=(2+\epsilon)/3$ is the Lorentz correction factor \cite{minavr2009electric}. It is also possible to detune the cavity rather than the ions by, for example, a piezoelectric effect \cite{goswami2018theory}.

It may be possible to avoid this detuning process between the entanglement generation step and the swapping step by making the cavity resonant with one transition (e.g., $\ket{\uparrow}$--$\ket{e}$) for entanglement generation and then choosing to use an off-resonant transition for the dissipative interaction required for entanglement swapping.

For the virtual exchange scheme, it is also necessary to tune the optical transitions of the ions into resonance with each other. In the case that we are able to address ions individually in space, this can be done by using, for example, the AC Stark effect. On the other hand, if individual addressing is not possible, we can use a large electric or magnetic field gradient to tune the transitions. 
The precision required for this resonance is determined by the gate time. Using equation (\ref{Lukin}) and the numerical values for fidelity and gate time given at the end of section \ref{exchangegatesection} for $C=9\times 10^4$, we estimate that the transitions should be resonant to within $2\pi\times0.8$ kHz for the deterministic scheme and within $2\pi\times1.6$ kHz when using perfect cavity monitoring. Although any amount of ion-ion detuning can cause infidelity, transitions further separated than this value will cause infidelity greater than the infidelity caused only by the finite cavity cooperativity. Note that detuning between transitions can actually be many times larger than their linewidths. This is because the value of ion-ion detuning $\Delta_w$ required to overtake the cooperativity-limited infidelity of $2\pi/\sqrt{C}$ is proportional to $\gamma C^{1/4}$.

However, after tuning the ions, to excite only one of the Er ions to the excited state, we still require the spatial resolution. For Er ions, which have long spontaneous emission time compared to the gate time, it might be possible to obviate this requirement by exciting one ion before bringing them in resonance. In this case, we should bring ions into resonance much faster than the gate time to keep the process adiabatic.

Efficient post-selection can enhance the fidelity of the gate for a given cavity cooperativity (or equivalently, reduce the cavity
cooperativity requirement for a given fidelity). This method is especially useful for RE ions, which typically couple to the cavity in the weak coupling regime. This is because the adiabatic condition needed to achieve a virtual photon interaction can be more easily violated for cavities near or within the strong coupling regime. Hence, in that regime, any gains in fidelity made by reducing the emitter-cavity detuning when post-selecting successful gates are offset by a decreased fidelity due to non-adiabatic phase evolution.


\textit{Electric dipole-dipole interaction}: To perform this gate, it is necessary to use ion transitions that are far detuned from the cavity. Otherwise, the excited state may decay due to the off-resonant Purcell effect before the gate is complete. This is apparent from Eq. (\ref{dipole}) where $F_\text{gate}\sim 1-T_\text{gate}\gamma/2$ implies that $\gamma\rightarrow \gamma + F_p\gamma_r$ would very quickly degrade the fidelity for a large Purcell factor $F_p$. If the cavity is resonant with $\ket{\uparrow}$--$\ket{e}$ for entanglement generation, then using the transition $\ket{\uparrow}$--$\ket{e^\prime}$ for the dipole-dipole interaction may place the transition $\ket{\downarrow}$--$\ket{e^\prime}$ close to the cavity resonance. Hence the difference between the Zeeman splitting of the ground and excited states $\delta_{eg}$ should be much larger than the cavity linewidth $\kappa$. Since $\delta_{eg}\simeq 2\pi\times100$ MHz, this implies it is necessary for the cavity quality factor to be larger than about $10^7$. For example, using  $F_p=5000$ and $\kappa=2\pi\times 16$ MHz corresponding to $Q=1.2\times 10^7$, the gate fidelity reduces to 0.951 due to an off-resonant Purcell factor of 32. 

Instead of requiring a far detuned transition, one could also actively detune the ions away from resonance with the cavity before performing the swapping gate using, for instance, the same methods mentioned for the virtual photon exchange scheme.

The electric dipole-dipole interaction  performs a deterministic gate that is very sensitive to the distance between the ions and requires them to be very close together (see Fig.\ref{fig:dipole-fidelity}). Hence, to perform the pulse sequence explained in Fig.~\ref{fig:electric-dipole}, it is still necessary to have either spatial or spectral resolution of the ions.

The dipole moment difference for $^{168}$Er$^{3+}$:Y$_2$SiO$_5$ is approximately $0.84\times10^{-31}$Cm \cite{Er-Eu}. If we assume the same value for $^{167}$Er$^{3+}$:Y$_2$SiO$_5$, then this gives an estimate for $\Delta\nu$ of $30$ and $0.03$ MHz for $r_{ij}=1$ and $10$ nm, respectively. 
These values are quite large compared to the magnetic dipole-dipole interaction between the ions. For $^{167}$Er with the magnetic moment of $-0.1618\ \mu_N$ \cite{zefoz-Er}, the magnetic dipole-dipole interaction is approximately $1.23$ and $0.001$ Hz for $r_{ij}=1$ and $10$ nm, respectively. As a result, the magnetic dipole-dipole interaction will not interfere with the electric dipole-dipole interaction.  

Performing the CNOT gate using the electric dipole-dipole interaction does not require a cavity itself; however, to generate entanglement and  to enhance the cycling transition (for the spin readout), the $\ket{\uparrow}$--$\ket{e}$ transition of the ions should be resonant with a cavity.

Unlike the virtual photon exchange scheme, the dipole-dipole scheme cannot take advantage of a high readout efficiency to improve fidelity by monitoring the cavity emission. This is because, in this scheme, the cavity does not mediate the interaction and so it is already necessary to minimize cavity emission by detuning it as far as possible. However, if a system can be optimized for a high collection efficiency of spontaneous emission directly from the Er ions without causing a Purcell enhancement, it may be possible to apply this same principle to the dipole-dipole gate. This type of collection enhancement could be implement using, for example, a combination of microfabricated solid-immersion lenses \cite{Hanson}, reflective coatings on one side of the substrate, and an objective with a large numerical aperture.




\subsection{State readout}\label{measurement}
In all of the explained schemes, a spin readout of each ion is required. To do so, we excite the $\ket{\uparrow}$--$\ket{e}$ transition of the Er ion and attempt to detect an emitted photon. The probability of emitting
a photon into the cavity mode (emission quantum efficiency) is $p=\eta_c\gamma_r F_p/(\gamma_{r} F_p+ \gamma)$.
Hence, for example, for $F_p=5000$ we expect $p=0.999\eta_c\simeq \eta_c$.
Even for $p=1$, the state measurement is limited by the efficiency of the single-photon detectors. Using superconducting detectors, the detection efficiency of more than $90 \%$ has been demonstrated \cite{detector1, detector2, detector3}. As shown in Sec. \ref{readout}, to improve the detection probability, we can repeatedly excite the ion in a cycling transition (through the $\ket{\uparrow}$--$\ket{e}$ transition) such that many photons will be emitted into the cavity and eventually at least one will be detected \cite{cycle, raha, Obrien}. Recently, it has been shown that a single $^{168}$Er ion doped Y$_2$SiO$_5$ crystal coupled to a silicon nanophotonic cavity can scatter more than 1200 photons using a single cycling transition \cite{raha}. Thus the probability that the cycle terminates during the small number of pulses needed to achieve a high fidelity readout is negligible. This also implies that a high-fidelity readout is possible even if the collection and detection efficiency is low by increasing the number of readout pulses. The consequence is that the readout time increases and, if comparable to the time needed to establish entanglement over an elementary link, may impact the distribution rate.


\section{conclusion and outlook}\label{ssec:conclusion}
The $^{167}$Er RE ion provides all of the desired features to implement the required elements of a quantum repeater. It has a nuclear spin coherence time within the one-second range, providing a natural long-lived quantum memory. It also has emission in the telecommunication wavelength window for low-loss long-distance transmission. Our proposed quantum repeater architecture utilizes a cavity-ion coupling to increase the spontaneous emission rate of the ion, improving the collection efficiency and single-photon indistinguishability. We discussed two different schemes to perform two qubit gates to achieve entanglement swapping within a repeater node. One can select the best scheme depending on the cavity characteristics and whether or not the ions are individually addressable in space or spectrum, or not at all. We have also shown how to improve the fidelity of a cavity-based virtual photon exchange entanglement swapping scheme by post-selecting successful gates on the absence of detected cavity emission. This post-selection approach could also be useful for other systems and gate schemes where cavity dissipation is the primary limitation for the fidelity.

We have shown that by using single ${^{167}}$Er ions, a higher entanglement distribution rate is possible compared to a hybrid single ${^{168}}$Er - ${^{151}}$Eu repeater scheme \cite{Er-Eu}. This entanglement distribution rate can even be further improved by multiplexing the protocol \cite{Er-Eu}. In terms of experimental feasibility, it is also easier to deal with a single species of ions rather than a doubly doped crystal.

Under certain conditions, a modified version of the Barrett-Kok scheme \cite{modified-BK}
can be used to perform a nearly deterministic swapping gate between nearby ions of a cavity.
This scheme does not require any individual addressing of ions or having them be close-lying. Instead, it needs ions to be in resonance with each other.
In this modified scheme, the ions are detuned from the cavity and both are excited to the state $\ket{e}$ simultaneously. The detection of one photon then projects one ion onto the state $\ket{\uparrow}$, but does not reveal which ion decayed. This generates an entangled state $\ket{e\uparrow}+ \ket{\uparrow e}$ between the ions. After the first photon detection, if both ion qubits are immediately flipped, and we wait for a second photon detection, the entangled state $\ket{\psi^+}$ is generated between the ions. Therefore, one can use this modified scheme to perform a CNOT
gate between ions in the same cavity. Because the excited-state lifetime of Er is so long, it should be possible to perform the feedback (spin flipping) fast enough to perform a nearly deterministic gate.

\section*{Acknowledgments}
FKA would like to thank P. Goldner, H. De Riedmatten, N. Lauk, S. Goswami, Y. Wu, J. Ji, S. Kumar and N. Sinclair for useful discussions. This work was supported by the Natural Sciences and Engineering Research Council of Canada (NSERC) through its  Discovery  Grant,  Canadian  Graduate  Scholarships and CREATE programs, and by Alberta Innovates Technology Futures (AITF).
 \section*{COMPETING INTERESTS}
The authors declare no competing interests.
	\bibliographystyle{apsrev4-1}
	\bibliography{ref}

\end{document}